\documentstyle[preprint,aps,epsfig]{revtex}

\begin{document}
\draft
\preprint{\vbox{
\hbox{IFT-P.002/99}
\hbox{January 1999}
}}
\title{ Scalar resonances in leptophilic  multi-Higgs models} 
\author{ J. C. Montero,  C. A. de S. Pires, and V. Pleitez }  
\address{
Instituto de F\'\i sica Te\'orica\\
Universidade Estadual Paulista\\
Rua Pamplona 145\\ 
01405-900-- S\~ao Paulo, SP\\Brazil} 
\date{\today}
\maketitle
\begin{abstract}
We show that the Higgs resonance can be amplified in a 3-3-1 model with 
a multi-Higgs ``leptophilic" scalar sector. This would allow the observation 
of the Higgs particle in muon colliders even for Higgs masses considerably higher 
than the ones expected to be seen in the electroweak standard model framework.
\end{abstract}
\vskip 0.7 cm
The study of the electroweak symmetry breaking is one of the main goals of
future colliders. The Higgs coupling to fermions are proportional to the 
fermion masses and hence the $s$-channel Higgs contribution to $e^+e^-\to 
f\bar{f}$ is highly suppressed in electron-positron colliders. However,
since muons are nearly 200 times heavier than electrons
it is usually considered in literature that the $s$-channel production of the 
Higgs boson in the $\mu^+\mu^-$ collider is one of the main features of muon 
colliders. The main process is also $\mu^+\mu^-\to f\bar{f}$~\cite{vb}.

We would like to point out that there exist a kind of models in which a scalar
multiplet couples only with leptons (not with quarks). 
If this is the case the vacuum expectation
value (VEV) of the neutral component of such multiplet can be naturally small
since it is the only responsible for generating the leptons masses.
An example of this kind of models are the multi-Higgs extensions of the 
electroweak standard model (ESM) with three Higgs doublets in such a way 
(say, by imposing an appropriate discrete symmetry) that one of them couples 
only with leptons in the fermion sector.  

There are models in which such a situation arises in a more
natural way than in the several doublets extensions of the ESM. 
These are models based on $SU(3)_C\otimes SU(3)_L\otimes U(1)_N$ gauge 
symmetry (3-3-1 model by short)~\cite{ppf}. In the minimal version of this
model a sextet $({\bf 6},0)$ is necessary for giving mass to the charged 
leptons and it does not couple to quarks indeed, so, its VEV does not have to 
be necessarily a large VEV, say it can be of the order of a few GeVs. 
A scalar sector having this particular characteristic will be called 
``leptophilic". Hence, leptophilic models can have a lepton-Higgs coupling stronger 
than the ESM one.
This characteristic will be the key for having a Higgs 
resonance--enhancing--effect in  processes involving leptons in the initial
and/or final state. Here we will consider a $\mu^+\mu^-$ collider to study the
Higgs-resonance in the reactions $\mu^+\mu^-\to W^+W^-,\, \mu^+\mu^-,\, b \bar b$,
($WW,\, \mu\mu, \, bb$, by short).

Throughout this work, for each process, we will refer to the pure Higgs contribution at
$\sqrt{s}=m_h$ as signal ($S=\sigma_h$), and to the non-Higgs ones as background ($B=\sigma_B$). 
Here we are also assuming 
that the pure 3-3-1 contributions to each reaction, {\it i.e.}, the ones that are not contained in
the ESM, are mediated by fields which are massive enough to make these contributions negligible. 
In this way the only difference between the 3-3-1 model and the ESM lies in the scalar sector so 
that the background will be the same for both models  for each process considered. The background
itself depends on the process: it is formed by the $\nu, \gamma$ and $Z$ contributions for $WW$ and
by the $\gamma$ and $Z$ ones for $\mu\mu$ and $bb$.  Once the ratio 
$N_S/\sqrt{N_B}={\cal L}^{\frac{1}{2}}_{\mu\mu}\, \sigma_h/\sqrt{\sigma_B}$ and ${\cal L}_{\mu\mu}$, 
the muon-collider luminosity, depends only on the machine, we  focus our attention on the 
model dependent dimensional ratio $\delta = S/\sqrt{B}=\sigma_h/\sqrt{\sigma_B}$ (cm). 
When necessary the expected First 
Muon Collider (FMC) luminosity is assumed: 
${\cal L}_{\mu\mu}=10^{33}\, \mbox{cm}^{-2}\, \mbox{s}^{-1}= 10 \, \mbox{fb}^{-1}\, \mbox{yr}^{-1}$. 
Although large polarization can enhance the ratio $S/B$\cite{kmp}, here we are not considering 
muon-beam polarization.

The total Higgs-width is model dependent, however, we are assuming that for the Higgs-mass range  
considered here  (100-200 GeV) no new 3-3-1 decay-modes
are allowed living the total $\Gamma_h$ equal to the one of the ESM.
In order to study the scalar resonance when comparing the 3-3-1 model with the ESM we will 
introduce the Higgs-signal-enhancing-factor (HSEF) $\mbox{f}$ defined for each process by  
$S^{331}/S^{\small ESM}$. 

The main goal of this paper is to show the possible existence of a Higgs-signal-enhancing effect  
due to the manifestation of the leptophilic quality of a given model, hence as we are not considering 
the experimental detection aspects in detail we assume that the detector efficiency is 1.

Although we will consider explicitly a 3-3-1 model,  we would like to stress
that our results will be valid in  multi-Higgs extensions of the ESM if
they are implemented in such a way that there is only a doublet coupling to 
leptons as we said above.


In the 3-3-1 model considered here the lepton mass term transforms as 
$({\bf1},{\bf3},0)\otimes({\bf1},{\bf3},0)=
({\bf1},{\bf3}^*,0)_A\oplus({\bf1},{\bf6},0)_S$, then we can introduce a 
triplet $\eta=(\eta^0,\eta^-_1,\eta^+_2)^T\sim({\bf1},{\bf3},0)$ and a 
symmetric sextet $S\sim({\bf1},{\bf6}^*,0)$. With the $\eta$-triplet only, 
one of the charged leptons remains massless and the two other are mass 
degenerate. Hence, the sextet $S$ at least has to be introduced in order to 
give arbitrary masses to all charged leptons. 
This scalar multiplet has the following charge attribution:
\begin{equation}
S=\left(
\begin{array}{ccc}
\sigma^0_1\; &\; h^+_2/\sqrt2 \;&\; h^-_1/\sqrt2\\
             h^+_2/\sqrt2 &\; H^{++}_1\; &\; \sigma^0_2/\sqrt2 \\ 
h^-_1/\sqrt2 \;&\; \sigma^0_2/\sqrt2 \;& \;H^{--}_2
\end{array}\right).
\label{2}
\end{equation}
In the model, in order to give mass to the quarks, it is necessary to have
two more triplets: $\rho=(\rho^+\,, \rho^0,\,\rho^{++})^T\sim({\bf1},\,
{\bf3},1)$ and $\chi=(\chi^-,\,\chi^{--},\,\chi^0)^T\sim({\bf1},\,{\bf3},\,-1)$.

Denoting the respective VEVs by $v_a$ ($a=\eta,\rho,\chi, \sigma_2$) and also 
assuming $v_{\sigma_1}=0$, 
in such a way that neutrinos remain massless (assuming that the total lepton number is 
conserved), we can write  
the masses of the vector bosons 
\begin{equation}
M_W^2=\frac{g^2}{4}(v_\eta^2 + v_\rho^2+v_{\sigma_2}^2 ),
\quad
M_Z^2\approx\frac{g^2}{4c^2_{W}}\left( v_\eta^2 + v_\rho^2+v_{\sigma_2}^2 
\right).
\label{wzmass}
\end{equation}

Notice that $v_\chi$ does not contribute to the vector boson masses (it only
contributes to the $Z$ boson mass but in such a way that it is proportional
to $v/v_\chi$, with $v$ being any of the other VEVs. Since the $v_\chi$ is the
VEV which is in control of the $SU(3)_L$ breaking, it is larger than the other
ones and we can neglect its contribution to $M_Z$. In this case the
physical neutral Higgs-scalars relevant at low energies are only those related to
$\eta,\rho$ and $S$ with the decoupling of the $\chi$ field~\cite{mdt}.
On the other hand, since the sextet $S$ couples only with leptons its VEV 
can be of the order of a few GeVs; its contribution to the vector boson masses 
can be smaller than the other VEVs contributions. 
Hereafter we will change the notation $v_\eta\to v_1$, $v_\rho\to v_2$ and
$v_{\sigma_2}\to v_3$ and $H^0_i$ ($i=1,2,3$) denote the real part of the scalar 
fields. These are related to the mass eigenstates $h^0_i$ by the orthogonal
matrix $O$: $H^0_i=O_{ij}h^0_j$. 

The interaction of the $W$ boson with the scalars is
\begin{equation}
{\cal L}_{hWW} = 
\frac{g^2}{2}\,v_i\,O_{ij}\,h^0_j\,W^+W^- ,
\label{6}
\end{equation}
while the Yukawa interaction in the lepton sector is given by 
\begin{equation}
{\cal L}^Y =
-\frac{m_l}{v_3}\,O_{3j}\,h^0_j\, \bar{l}\, l,
\label{10}
\end{equation}
since only the sextet ($\sigma_2^0=O_{3j}h_j$) couples with them. Notice that this scalar interaction 
with leptons are not necessarily negligible since $v_3$ can be of the 
order of some GeVs as commented earlier.

We will study the $\mu^+ \mu^- \rightarrow W^+W^-$  process in some detail
and for the other ones we will omit some of the technicalities for shortness.
There are four contributions to the $WW$ process  
being three of them in the $s$-channel with $Z^0$-, $A$(photon)- and $h^0$'s 
exchanges and one in the $t$-channel with a neutrino exchange.

The amplitude ${\cal M}_{WW}$ is:
\begin{equation}
{\cal M}_{WW}={\cal M}^\nu+{\cal M}^A+{\cal M}^Z+{\cal M}^{h_j}.
\label{15}
\end{equation}
In the 3-3-1 model all the $W$-couplings have the same form as in the 
standard electroweak model. The neutrino $M^\nu$ and the vector boson 
contributions ($M^\gamma,M^Z$) are exactly the same as in the ESM and are given by
\begin{equation}
{\cal M}^\nu=\frac{-4iG_F m_W^2}{\sqrt{2}t}\bar v(p_2,r_2)H_R 
\gamma^\rho \gamma_\mu k_1^\mu \gamma^\sigma u(p_1,r_1) \epsilon^\rho 
\epsilon^\sigma,
\label{17}
\end{equation}
\begin{eqnarray}
{\cal M}^\gamma &=&\frac{8iG_F m_W^2\sin^2\theta_W}{\sqrt{2}s}\bar 
v(p_2,r_2)\gamma_\nu u(p_1,r_1) \epsilon^\rho \epsilon^\sigma
\cdot [(p_4-p_3)^\nu g^{\rho \sigma}\nonumber \\ &&\mbox{}
-(k_2+p_4)^\sigma g^{\rho \nu}+
(k_2+p_3)^\rho g^{\nu \sigma}],
\label{18}
\end{eqnarray}
\begin{eqnarray}
{\cal M}^Z&=&\frac{-4iG_F m_Z m_W^2 \cos\theta_W(s-m_Z^2-im_Z 
\Gamma_Z)}{\sqrt{2}[(s-m_Z^2)^2+m_Z^2 \Gamma_Z^2]}\nonumber \\
&&\cdot\bar v(p_2,r_2)[R H_R +L H_L]\gamma_\mu u(p_1,r_1)(g_{\mu 
\nu}-\frac{k_{2\mu}k_{2\nu}}{m_Z^2}) \epsilon^\rho \epsilon^\sigma\nonumber \\
&&\cdot[(p_4-p_3)^\nu g^{\rho \sigma}-(k_2+p_4)^\sigma g^{\rho \nu} +(k_2+p_3)^\rho 
g^{\nu \sigma}],
\label{19}
\end{eqnarray}
where $H_{R,L}=\frac{1}{2}(1\pm\gamma^5)$; $R=(2/3)\sin^2\theta_W$, 
$L=-1+(2/3)\sin^2\theta_W$;  $p_1,p_2,p_3,p_4$ are the momenta 
of the electron, positron, $W^-$, and $W^+$, respectively; $k_2=p_3+p_4$ 
is the momentum of the photon and $Z$. 

However, the  scalar contributions are different in 3-3-1 models. Each 
s-channel exchange of a physical Higgs $h_j$ gives the contribution:
\begin{equation}
{\cal M}^{h_j}=\frac{4iG_F m_\mu m_W^2 
O_{3j}}{\sqrt{2}\, v_3\left((s-m_j^2)^2+m_j^2\Gamma^2_j 
\right)}\sum_iO_{ij}v_i\,\left((s-m_j^2)-im_j\,\Gamma_j \right)
\,\bar v(p_2,r_2)u(p_1,r_1)g_{\rho \sigma} 
\epsilon^\rho \epsilon^\sigma,
\label{16}
\end{equation}
where $m_j$ and $\Gamma_j$ are 
the mass and total width of the scalar $j$ (with no sum in $j$).

In the ESM a scalar resonance can only be revealed
in a $\mu$--collider for a relatively light Higgs. This can be understood by 
the following simple reasoning.
At the Higgs pole ($\sqrt{s} = m_h$) the squared-Higgs-amplitude
is proportional to $1/m_h^2\Gamma_h^2$ being the total Higgs width 
$\Gamma_h$ a fast growing function of the Higgs mass: it goes from
$\sim 3.1\times 10^{-3}$ GeV for $m_h \sim 110$ GeV to $\sim 1.6$ GeV for 
$m_h \sim 200$ GeV. Hence the total Higgs contribution is suppressed for
relatively large values of the Higgs mass. This argument is also valid for 
$\mu\mu$ and $bb$.

On the other hand for a 3-3-1 model, as it can be seen from Eq.(\ref{16}), 
the squared-Higgs-amplitude at $\sqrt{s}=m_{j}$ is proportional to
\begin{equation}
\frac{1}{m_{j}^2 \Gamma_{j}^2} \left[
\frac{O_{3j}^2}{v_3^2} \left(O_{1j}v_1+O_{2j}v_2+O_{3j}v_3\right)^2
\right].
\label{r331}
\end{equation}
Where the mixing--angles must obey the unitary condition $\sum_j O_{3j}^2=1$ 
and the VEVs are related by the constraint $v_1^2+v_2^2+v_3^2=v_{\small ESM}^2=(246\;
\mbox{GeV})^2$.
As we said before once the VEV $v_3$  is only 
needed to give mass to the massive leptons, it
can take small values so that the VEV-mixing-angles relation 
between the square brackets above can enhance the total Higgs contribution
in such way that it can be still significant even for large values of the 
Higgs mass.

In this case the HSEF is given by
\begin{equation}
\mbox{f}_{WW}=
\frac{O_{33}^2}{v_3^2} \left(O_{13}v_1+O_{23}v_2+O_{33}v_3\right)^2.
\label{fww}
\end{equation}
The behaviour of $\mbox{f}_{WW}$ is showed in Fig.~(1) as a function of $v_3$ when a choice for 
the others parameters is made.
In Fig.(2) we show S and B for 
the $WW$ process for the ESM and the 3-3-1 model. There we can see that due to the HSEF,  
S is considerably enlarged for the 3-3-1 model providing a better 
ratio $S/\sqrt{B}$ when comparing with the ESM.  We have built 
these figures by considering the s-channel exchange of a single Higgs in the 
3-3-1 amplitudes. Here we have taken $j=3$, however, in order to unify the notation we use 
$m_3=m_h$ and $\Gamma_3=\Gamma_h$. For very large values of $\sqrt{s}$  we certainly must 
add all others scalar contributions of the 3-3-1 model (and the boson 
$Z^\prime$ as well) in order to ensure unitarity. However, as we said before, for the values of 
$\sqrt{s}$ we are considering here we are assuming that the other scalars 
$h_1,h_2$ (and also the $Z^\prime$) are massive enough to not affect this 
picture.  Here we have considered only the case of a real $W$-pair production 
to illustrate that even for a relatively massive scalar a resonance can occur
in the $WW$ process.
However, the enhancing factor that we have pointed out here will also occur 
with a virtual pair production since it is introduced by a fundamental vertex
of the model.


Next we consider the $\mu^+\mu^-\to \mu^+\mu^-$ process. 
As it is well known the Higgs signal is very small in the ESM framework provided 
that each vertex introduces 
a $m_\mu/v_{\small ESM}$ factor, where $m_\mu$ is the muon mass and $v_{\small ESM}=246$ GeV. 
For this process we have the contribution of the $t$ and the 
$s$--channel, however, the last one is largely dominant at the resonance. 
Hence the  main $s$-channel Higgs contribution to $S^{\small ESM}$ 
is proportional to $(m_\mu/v_{\small ESM})^4/(m_{3}^2 \Gamma_{3}^2)$ at the Higgs peak. 
Since $m_\mu/v_{\small ESM}$ is an additional suppression factor and   
 following the same argument we used before, 
a ESM Higgs-resonance are not expected to be seen in this process. On the other hand in the 3-3-1 model, due 
to the Yukawa coupling given by Eq.~(\ref{10}), the Higgs signal $S^{331}$ is 
proportional to $(O_{33})^4(m_\mu/v_3)^4/(m_h^2\Gamma_{h}^2)$ at the resonance. 
As we said early, once $v_3$ can be chosen relatively small ($10-50$ GeV), the factor 
$(O_{33})^4(m_\mu/v_3)^4$ can considerably enhance the Higgs signal when comparing with the 
ESM one. In fact the HSEF $\mbox{f}_{\mu\mu}$ is  given by
\begin{equation}
\mbox{f}_{\mu\mu}=(O_{33})^4(v_{\small ESM}/v_3)^4,
\label{fmm}
\end{equation}
and this quantity can vary from $10^2$ for $v_3=50$ GeV to $10^5$ for 
$v_3=10$ GeV assuming that $O_{33} \sim {\cal O}(1)$ as showed in Fig.~(1).
In Fig.~(3) we show  $S$ and $B$ as a function of $\sqrt{s}=m_h$ and 
$\sqrt{s}$ respectively. We do not quote $\sigma^{\small ESM}_h$ because it is negligible. There 
we can see that for relatively light Higgs $\sigma^{331}_h$ dominates the cross section, and that 
for $m_h > 135$ GeV  $S^{331}$ goes under the $B$ and falls 
rapidly to zero, however  the resulting HSEF  
is still enough to allow for a Higgs-resonance detection in the range $m_h \sim 150-160$ GeV  
in the $\mu^+\mu^-\to \mu^+\mu^-$ process. Although that enhancing--factor 
can provide a very pronounced peak in the total cross section at the Higgs resonance for 
relatively light Higgs, we must remember that for this sort of masses the Higgs-width is very 
small ant this will require a very high-resolution energy scan. As it was shown in Fig.~(2), 
this is not the case for the $\mu^+\mu^- \to W^+W^-$ process:  $S^{331}$ remains appreciable 
even for $m_h > 170$ GeV. In this case a resonance detection is much easier since for masses in 
this range the Higgs-width is considerably large and so the energy-resolution requirements can be 
less stringent.


Finally we will examine the $\mu^+\mu^-\to b {\bar b}$ process.
In the ESM the $b{\bar b}$ final state  
considerably increases the $S^{\small ESM}$ due to the $b$-quark mass as it is shown in Fig.~(4). 
There we can see that, differently from the $\mu\mu$ process,  the Higgs signal 
dominates over the background up to $m_h < 145$ GeV. Although the total cross section for $b{\bar b}$
production is much lower than for $\mu^+\mu^-$, the higher ratio $S/\sqrt{B}$ provided by  $bb$ 
make this one the process to be studied at muon colliders in order to detect a relatively light 
Higgs-resonance \cite{vb}.

In the 3-3-1 model the couplings lepton--scalar and quark--scalar are different. Once $\sigma_2^0$ 
couples only with leptons, the lepton--scalar vertex is provided by Eq.~(\ref{10}),  while for $b$-like 
quarks the main flavor-conserving quark--scalar interactions are given by
\begin{equation}
{\cal L}^Y_{d-like} =
-\frac{m_b}{v_\eta}\,O_{\eta j}\,h^0_j \,\bar{b}\, b,
\label{hbb}
\end{equation}

Hence, the pure $s$-channel Higgs contribution to $b{\bar b}$ at the 
resonance is proportional to 
\begin{equation}
\frac{1}{m_{h}^2 \Gamma_{h}^2}(O_{33}O_{13})^2 
\left( \frac{m_\mu m_b}{v_3 v_1} \right)^2,
\label{s331bb}
\end{equation} 
where we have used the previous redefinition: $v_\eta\to v_1$, $v_\rho\to v_2$ and
$v_{\sigma_2}\to v_3$. In this case, for $\sqrt{s}=m_h$, the HSEF $\mbox{f}_{b{\bar b}}$ 
is given by  
\begin{equation}
\mbox{f}_{b{\bar b}}=(O_{33}O_{13})^2\left( \frac{v_{\small ESM}^2}{v_3 v_1} \right)^2.
\label{fbb}
\end{equation} 
As before this factor allows the possibility of having an enhancement in 
the pure Higgs amplitude as showed in Fig.~(1).

In Fig.~(5) we show the quantity $\delta$ defined above for the three processes and the 
two model here considered.  We omit $\delta_{\mu\mu}$ for 
the ESM because it is negligible. From Fig.~(5) we can see that in the 
ESM framework, and assuming that $N_S/\sqrt{N_B} \ge 5$ is necessary for 
detection, the $bb$ process is able to detect Higgs-resonances for $m_h$ 
up to $\sim$162 GeV. Beyond this Higgs-mass, only the $WW$ process provide a 
$\delta_{WW}$ large enough to allow for detections  for $m_h$ up to $\sim$180 GeV. 
In the same way, Fig.~(5) also shows that in the 3-3-1 model framework the 
$\mu\mu$ process is sensitive to detect Higgs-resonances for $m_h$ up to $\sim$167 GeV,
the $bb$ process up to $\sim$180 GeV, and that the $WW$ is the only process able to 
investigate Higgs-resonances for $m_h$ in the heavy-mass range $m_h > 200$ GeV.      
Here we are using the FMC luminosity 
${\cal L}_{\mu\mu}=10^4\, \mbox{pb}^{-1}\, \mbox{yr}^{-1}$ 
so that in order to be compatible with the $N_S/\sqrt{N_B} \ge 5$ requirement, we 
must have $\delta \ge 5\times 10^{-2} \, \mbox{pb}^{-1/2}$. This limit line $\delta_l$ is also 
showed in Fig.~(5). The values given above concerning the 3-3-1 model should be taken 
as indicative only once they depend on a set of parameters that are a priori free. 
In Tab.~1 we summarize the above results  given $S$ and $B$, and the corresponding 
$N_S$ and $N_B$, and  accuracies, for these representative Higgs-masses.

The existence of a Higgs-resonance far above the ESM-limit will indicate 
at least that that scalar no longer belongs to the minimal ESM and that 
new physics must be brought in. 
Although this resonance may exist in a multi-Higgs $S(U2)_L\otimes U(1)_Y$ 
model, this is only possible by extending the symmetry with an discrete
one. On the other hand, in the 3-3-1 model the fact that the sextet couples 
only to leptons is just a consequence of the representation content in the 
lepton sector in the minimal model. Hence, the scalar resonance  seems more 
natural in the last model than in the multi-Higgs extensions of the 
electroweak standard model.

\newpage
\acknowledgments 
This work was supported by Funda\c{c}\~ao de Amparo \`a Pesquisa
do Estado de S\~ao Paulo (FAPESP), Conselho Nacional de 
Ci\^encia e Tecnologia (CNPq) and by Programa de Apoio a
N\'ucleos de Excel\^encia (PRONEX). One of us (CP) would like to thank
Coordenadoria de Aperfei\c coamento de Pessoal de N\'\i vel Superior (CAPES) 
for financial support.

\narrowtext

\newpage

\begin{center}
{\bf Figure Captions}
\end{center}
\vskip .5cm

\noindent {\bf Fig.~1} The HSEF $\mbox{f}_{WW}$ (continuous line),  $\mbox{f}_{\mu\mu}$ 
(dashed line ), and $\mbox{f}_{b{\bar  b}}$ (doted line) as a function of $v_3$ for
$O_{31}=0.18, O_{32}=0.085$, $O_{33}=0.98$, and for  
$v_1=v_2=\sqrt{(v_{\small ESM}^2-v_3^2)/2}$.

\vskip .5cm
\noindent {\bf Fig.~2} S for the 3-3-1 model (solid line) and for the ESM 
(dashed line) as a function of $\sqrt{s}=m_h$, and B (doted line) as a function of 
$\sqrt{s}$ for the $WW$ process with $v_1=v_2=173.8$ GeV, $v_3=10$ GeV,  
$O_{31}=0.18, O_{32}=0.085$, and $O_{33}=0.98$.

\vskip .5cm
\noindent {\bf Fig.~3} S for the 3-3-1 model (solid line) as a function of 
$\sqrt{s}=m_h$ and B (doted line) as a function of $\sqrt{s}$ for the 
$\mu\mu$ process with $v_3=10$ GeV.

\vskip .5cm
\noindent {\bf Fig.~4} S for the 3-3-1 model (continuous line), and for the 
ESM (dashed line) as a function of $\sqrt{s}=m_h$ and B (doted line) as 
a function of $\sqrt{s}$ for the $bb$ process with  $v_3=10$ GeV, $v_1=173.8$ 
GeV, $O_{33}=0.98$, and $O_{13}=0.18$.

\vskip .5cm
\noindent {\bf Fig.~5} The quantity $\delta=S/\sqrt{B}$ as a function of 
$m_h$ for the $WW,\, \mu\mu$, and $bb$ processes for the  3-3-1 model 
and the ESM, as indicated on the lines, for the same parameters we have 
used in Figs.~2--4. The limit-line $delta_l=0.05\, pb^{-1/2}$ for 
$N_S/\sqrt{N_B}=5$ for ${\cal L}_{\mu\mu}=10^4\,pb^{-1}yr^{-1}$.

\newpage
\begin{figure*}
\mbox{\epsfxsize=430pt \epsffile{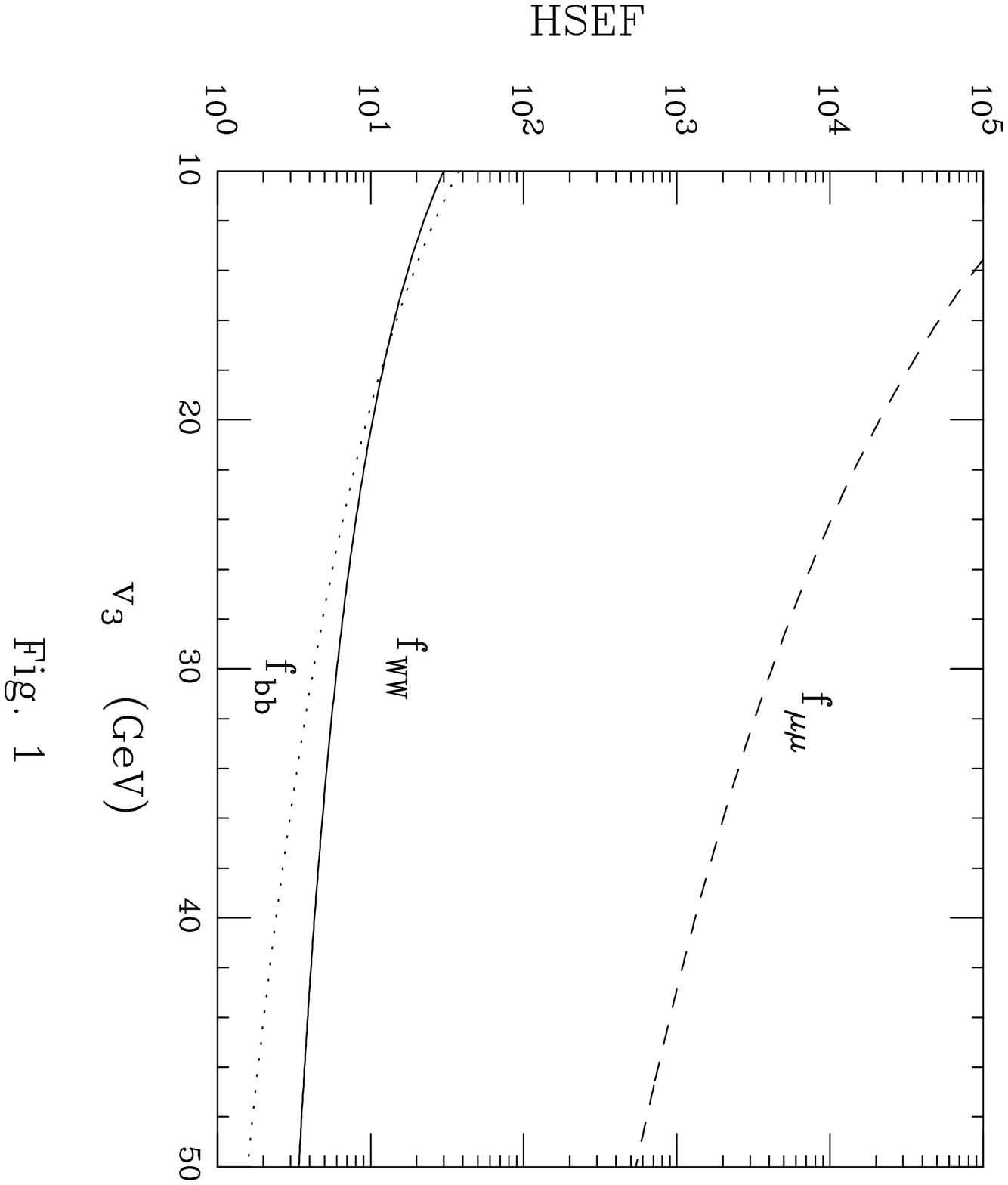}}
\end{figure*}

\newpage
\begin{figure*}
\mbox{\epsfxsize=430pt \epsffile{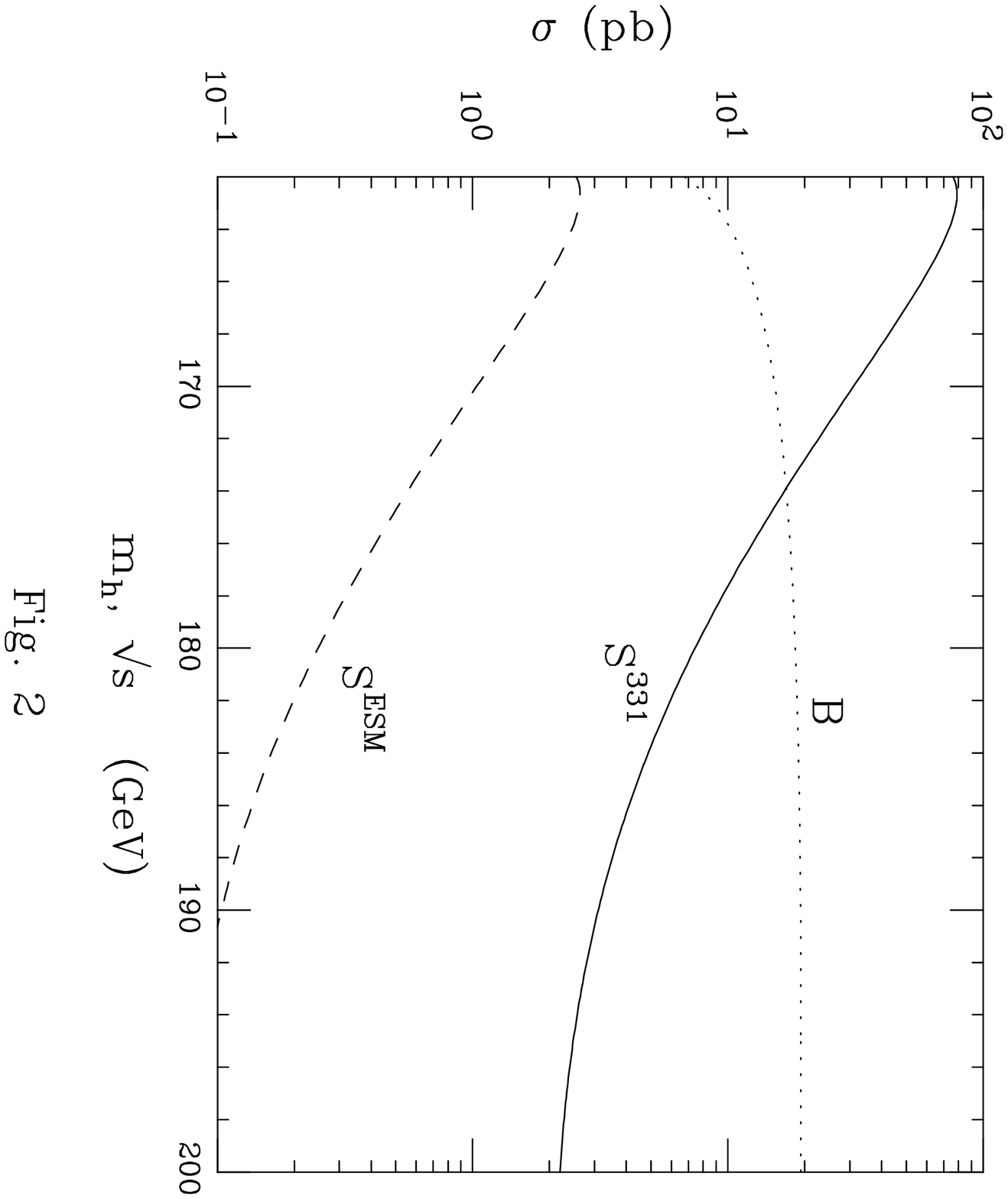}}
\end{figure*}

\newpage
\begin{figure*}
\mbox{\epsfxsize=430pt \epsffile{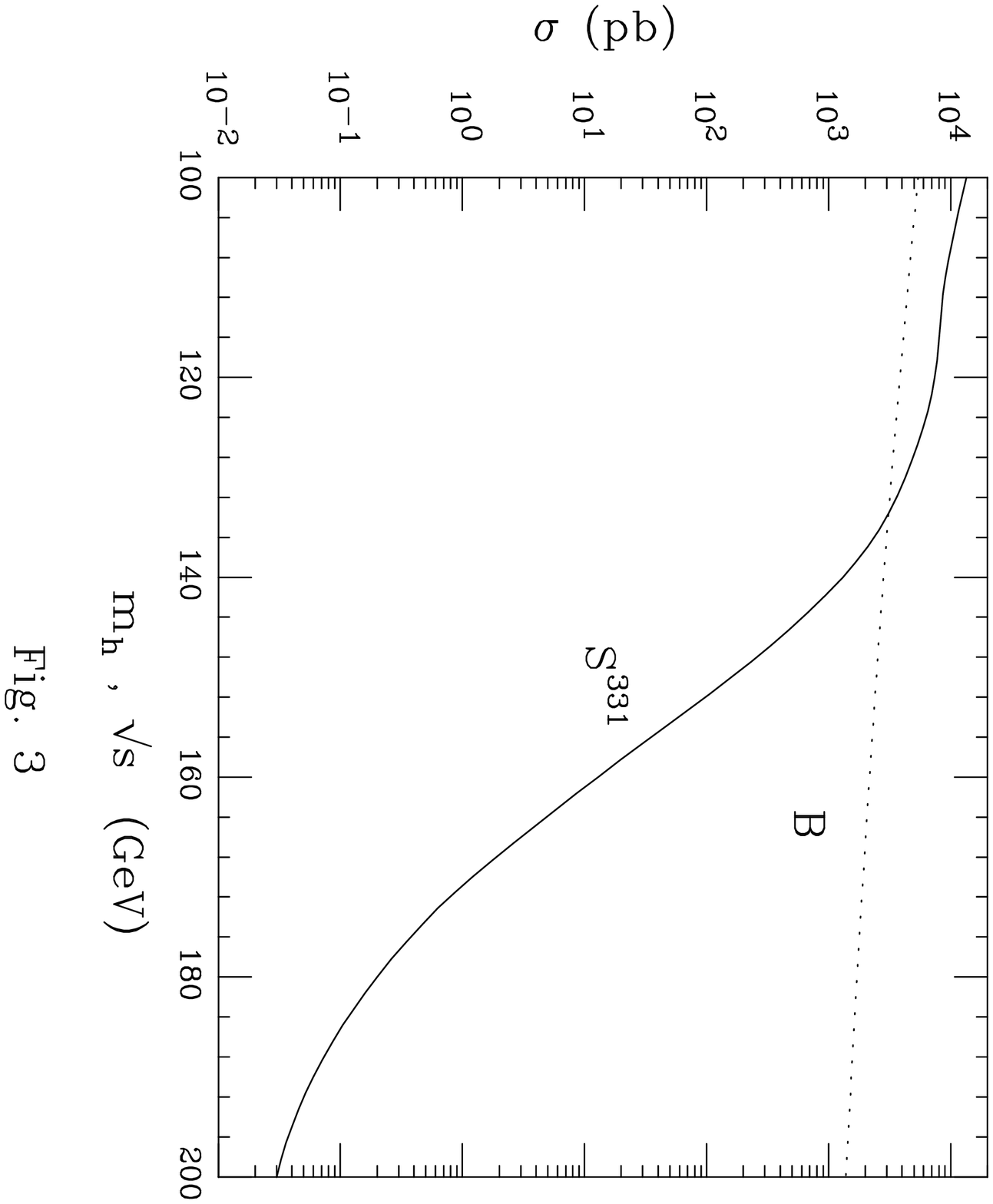}}
\end{figure*}

\newpage
\begin{figure*}
\mbox{\epsfxsize=430pt \epsffile{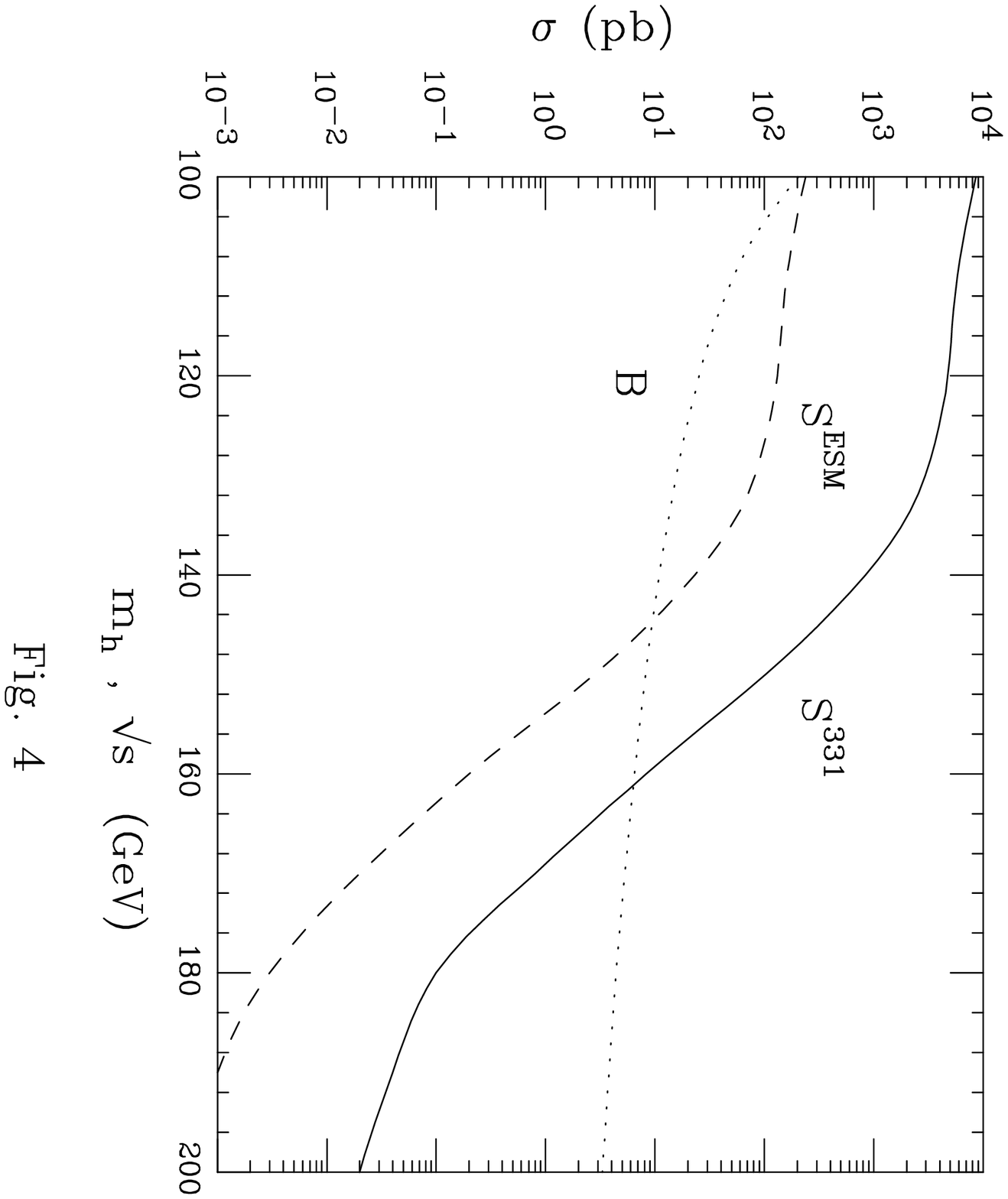}}
\end{figure*}

\newpage
\begin{figure*}
\mbox{\epsfxsize=430pt \epsffile{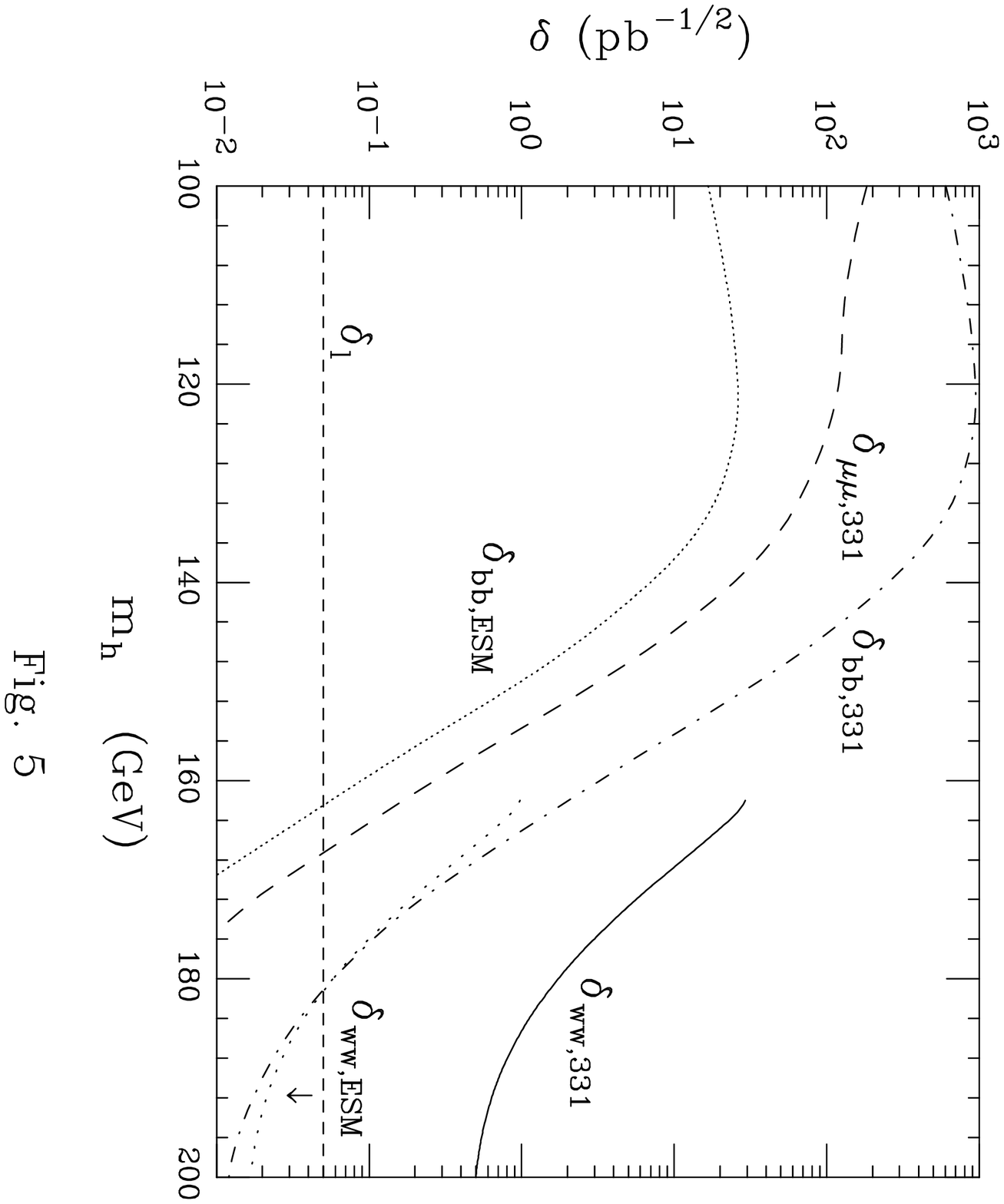}}
\end{figure*}

\newpage
\begin{table}
\caption{The quantities $S$, $B$, $N_S$, $N_B$, and the accuracies $N_S/\sqrt{N_B}$ and $\sqrt{N_S+N_B}/N_S$ 
for $\sqrt{s}=m_h$, and for the FMC luminosity.}
\begin{center}
\begin{tabular}{l|c|c|c|c|c|c|c} 
process&$m_h$ (GeV)&$S$ (pb)&$N_S$ (events)&$B$ (pb)& $N_B$ (events)&$N_S/\sqrt{N_B}$&$\sqrt{N_S+N_B}/N_S$   \\ \hline
bb-ESM &162& 0.138   & 1380        & 6.219  & 62190         & 5.5            & 0.18                  \\ \hline
WW-ESM &180& 0.265   & 2650        & 19.364 & 193640        & 6.0            & 0.16                  \\ \hline
$\mu\mu$-331&167& 2.334 & 23340    & 1984.943& 19849430     & 5.2            & 0.19                 \\ \hline
bb-331 &180& 0.123   & 1230        & 4.431   & 44310        & 5.8            & 0.17                  \\ \hline
WW-331 &200& 1.153   & 11530       & 19.263  & 192630       & 26.3           & 0.04                  \\ 
\end{tabular}
\end{center}
\end{table}

\end{document}